\documentclass[shortnote,twocolumn]{jpsj2}
\usepackage{graphicx}

\def\runtitle{Numerical Studies of Fano Resonance}
\def\runauthor{A.\ \textsc{Ueda}, I.\ \textsc{Baba},
               K.\ \textsc{Suzuki} and M.\ \textsc{Eto}}

\title{%
Numerical Studies of Fano Resonance in Quantum dots Embedded in AB Rings
}

\author{%
Akiko \textsc{Ueda}, Ikuo \textsc{Baba},
Keita \textsc{Suzuki} and Mikio \textsc{Eto}\thanks{Corresponding author.
E-mail: eto@rk.phys.keio.ac.jp.}
}

\inst{%
Faculty of Science and Technology, Keio University,
3-14-1 Hiyoshi, Kohoku-ku, Yokohama 223-8522, Japan
}

\recdate{\today}

\abst
{
}

\kword{%
quantum dot, AB ring, Fano resonance, phase lapse, dephasing,
resonant tunneling, Aharonov-Bohm effect
}

\begin{document}
\sloppy
\maketitle


In Aharonov-Bohm (AB) rings with a quantum dot embedded in one arm,
the Fano resonance has been observed recently.\cite{Kobayashi}
In previous works, the AB oscillation was reported in such systems,
as a function of magnetic flux enclosed in the ring.\cite{Yacoby,Schuster}
The AB oscillation is caused by the interference between the
electronic wave passing through one arm and that through the other,
as in a double-slit interferometer.
Compared with the AB oscillation, the higher-order
interference effect between the two paths (a discrete level
$\varepsilon_0$ in the quantum dot and a continuum of states in the
other arm) results in the Fano resonance.\cite{Fano}
The conductance shows an asymmetric shape of the resonance,
$G \propto (e+q)^2/(e^2+1)$, with $e=(\varepsilon-\varepsilon_0)/\Gamma$,
where $\varepsilon$ is the energy of an incident electron (Fermi energy)
and $\Gamma$ is the line broadening.

The Fano resonance observed in the quantum-dot systems shows unique
characters.\cite{Kobayashi} (i) With increasing magnetic flux, an
asymmetric resonance-shape with negative $q$ changes to that with
positive $q$, via a symmetric resonance-shape.
(ii) Around the center of resonance,
the ``phase'' seems to change continuously in two-terminal set-ups,
although the phase jumps by $\pi$ in the AB oscillation observed by
the two-terminal measurement.\cite{Yacoby} (iii)
The Fano resonance is observable only when the high coherence is kept
in the whole system.
These properties have not been fully understood in spite of several
theoretical works including the electron-electron
interaction in quantum dots.\cite{Bulka,Hofstetter,Kim}
In this short note, we examine two models of non-interacting electrons.
The merit of model (a) in Fig.\ 1 is to obtain an exact expression
for the resonance as a function of magnetic flux. The calculated results
clearly explain the experimental findings (i) and (ii).
Regarding (iii), we consider the dephasing effect on the Fano resonance 
in model (b) by numerical calculations.

\begin{figure} 
\begin{center}
\includegraphics[width=6cm]{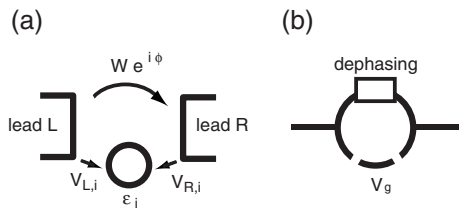}
\caption{Two models we study for the transport properties of
quantum dots embedded in AB rings.}
\end{center}
\end{figure}

In model (a), the magnetic flux enclosed in the ring is
taken into account by the phase factor $\phi$ of the direct
tunnel-coupling, $W$, between leads L and R. We obtain the conductance
in an analytical form. In the case of single level ($\varepsilon_0$)
in the quantum dot, the conductance is written as an extended
Fano form,\cite{com1}
\begin{equation}
G = \frac{2e^2}{h} \frac{4\lambda^2}{(1+\lambda^2)^2} 
\frac{|e+q|^2}{e^2+1},
\end{equation}
with a complex parameter
\begin{equation}
q=\frac{V_{\rm L}V_{\rm R}}{V_{\rm L}^2+V_{\rm R}^2}
  \frac{1+\lambda^2}{\lambda}
\left[ \frac{1-\lambda^2}{1+\lambda^2}
\cos\phi - i \sin\phi \right].
\end{equation}
Here, $\lambda=\pi\nu W$ and
$e=(\varepsilon-\tilde{\varepsilon_0}(\phi))/\tilde{\Gamma}$, where
$\nu$ is the density of states in the leads,
$\tilde{\varepsilon_0}(\phi)=\varepsilon_0-\frac{\lambda}{1+\lambda^2}
(2\pi\nu V_{\rm L}V_{\rm R})\cos\phi$ and
$\tilde{\Gamma}=\Gamma/(1+\lambda^2)$ with $\Gamma$ being
$\pi\nu(V_{\rm L}^2+V_{\rm R}^2)$.
Parameter $\lambda$ characterizes the strength of direct transmission
between the leads ($0\le \lambda \le 1$).\cite{com2}
Figure 2 (upper panel) shows $G$ as a function of
$\varepsilon$ when $\lambda=0.3$.
With increasing $\phi$ (magnetic flux in the ring),
an asymmetric line-shape of $G$ (Re$q >0$ at $\phi=0$) changes to
symmetric (Re$q =0$ at $\phi=\pi/2$) and
to asymmetric (Re$q <0$ at $\phi=\pi$),
in accordance with the experimental results mentioned above.

\begin{figure} 
\begin{center}
\includegraphics[width=6cm]{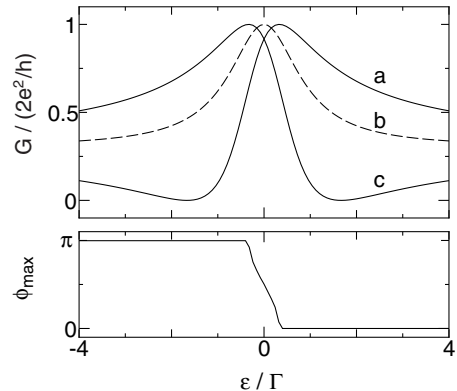}
\caption{The conductance (upper panel) and phase $\phi_{\rm max}$
(lower panel), as functions of the Fermi energy $\varepsilon$,
in model (a) with single level ($\varepsilon_0=0$). $\lambda=0.3$.
In the upper panel, the magnetic flux is $a$, $\phi=0$;
$b$, $\phi=\pi/2$; and $c$, $\phi=\pi$.}
\end{center}
\end{figure}

We define ``phase'' of the resonance as $\phi_{\rm max}$,
the value of $\phi$ when $G(\phi)$ is maximal.
As shown in Fig.\ 2 (lower panel), $\phi_{\rm max}$
changes from $\pi$ to $0$
rapidly but continuously with increasing $\varepsilon$. Note that
this does not contradict the Onsager's relation, $G(\phi)=G(-\phi)$, in
two-terminal situations. In accordance with the relation,
$G$ in eq.\ (1) is a function of $\cos\phi$; $G=F(\cos\phi)$.
At the maximum of $G$, $F'(\cos\phi)\sin\phi=0$. When the equation
of $F'(\cos\phi)=0$ has solutions, $0<\phi_{\rm max}<\pi$. Otherwise
$\phi_{\rm max}=0$ or $\pi$, which is always the case of AB
oscillation.\cite{Yacoby}

For more than one level in model (a), the conductance can be expressed
in terms of the Green's functions (Appendix). The conductance
$G(\phi=0)$ and $\phi_{\rm max}$ are shown in Fig.\ 3,
in the case of two levels with
$V_{{\rm L},1}=V_{{\rm R},1}=V_{{\rm L},2}=\pm V_{{\rm R},2}$.
When $V_{{\rm L},1}/V_{{\rm R},1}$ and $V_{{\rm L},2}/V_{{\rm R},2}$
are in phase, two Fano resonances are asymmetric in the same direction.
$\phi_{\rm max}$ changes continuously by $-\pi$ at both the resonances.
Between them, $\phi_{\rm max}$ jumps by $\pi$ (phase lapse).
When $V_{{\rm L},1}/V_{{\rm R},1}$ and $V_{{\rm L},2}/V_{{\rm R},2}$
are out-of-phase, the Fano resonances are asymmetric
in the opposite direction. $\phi_{\rm max}$ changes by $-\pi$ at one
resonance and by $\pi$ at the other resonance.
In this case, no phase lapse takes place.

\begin{figure} 
\begin{center}
\includegraphics[width=7cm]{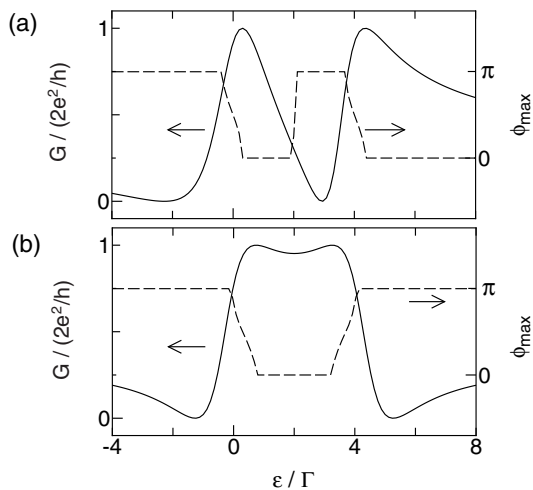}
\caption{The conductance $G(\phi=0)$ and phase $\phi_{\rm max}$,
as functions of the Fermi energy $\varepsilon$, in model (a)
with two levels ($\varepsilon_1=0$, $\varepsilon_2=4\Gamma$).
$\lambda=0.3$.
(a) $V_{{\rm L},1}=V_{{\rm R},1}=V_{{\rm L},2}=V_{{\rm R},2}$,
whereas (b)
$V_{{\rm L},1}=V_{{\rm R},1}=V_{{\rm L},2}=-V_{{\rm R},2}$.}
\end{center}
\end{figure}

Model (b) consists of one-dimensional perfect wires.
The quantum dot is formed between two $\delta$-functions
($V_{\rm L}\delta(x-x_{\rm L})$, $V_{\rm R}\delta(x-x_{\rm R})$),
the electrochemical potential of which is controlled by $V_{\rm g}$. 
The ring is 1000\AA \ in radius. The dot size is
$L=x_{\rm R}-x_{\rm L}=1000$\AA \ and
$V_{\rm L, R}=\hbar^2/(2m^*L)\times 10$, $15$ ($m^*=0.067m$).
The dephasing effect is taken into account on the upper arm
phenomenologically; a part ($\alpha$) of the current loses the phase
information in virtual electrodes.\cite{Buttiker}
We evaluate the conductance numerically
using the $S$-matrix method. The calculated results with $\alpha=1/2$ are
given in Fig.\ 4. The conductance as a function of $eV_{\rm g}$ shows
an asymmetric shape of resonance (solid line) or symmetric one (dotted line),
depending on the magnetic flux enclosed in the ring.\cite{com3}
The amplitude of the resonance
decreases as the dephasing effect is enhanced ($\alpha$ is larger).
The Fano resonance is clearly observed even with $\alpha > 1/2$,
which indicates the robustness of the resonance. This seems
against the experimental results and should be attributable to the simplicity
of the model, {\it e.g}.\ disregarding the thickness of the wires.

\begin{figure} 
\begin{center}
\includegraphics[width=7cm]{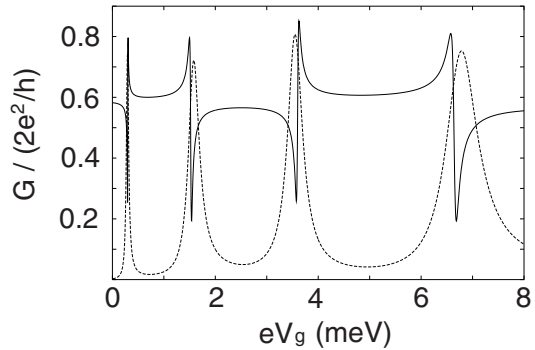}
\caption{The conductance as a function of $eV_{\rm g}$ in model (b).
The energy levels are shifted by $-eV_{\rm g}$ in the quantum dot,
whereas the Fermi level is fixed.
The dephasing rate is $\alpha=1/2$. The magnetic flux
in the ring is $\Phi/(\hbar/e)=0$ (solid line) and $\pi/2$ (dotted line).}
\end{center}
\end{figure}

\appendix
\section{Calculations of conductance in \\ model (a) with multilevel}
The Green's functions
$G_{ij}(\varepsilon)$, with $i$ and $j$ being level indices in the quantum
dot, are given by
\begin{eqnarray}
\sum_j \left[ (\varepsilon-\varepsilon_i) \delta_{ij}-\Sigma_{ij}
\right]G_{jk}(\varepsilon)=\delta_{ik}, \\
\Sigma_{ij}=\frac{1}{1+\lambda^2}
\Bigl[ -i\pi\nu (V_{{\rm L},i}V_{{\rm L},j}^*+V_{{\rm R},i}V_{{\rm R},j}^*)
\nonumber \\
-\lambda\pi\nu (e^{-i\phi}V_{{\rm L},i}V_{{\rm R},j}^*+
e^{i\phi}V_{{\rm R},i}V_{{\rm L},j}^*) \Bigr].
\end{eqnarray}
The $T$-matrix between states in lead R and those in lead L is
\begin{equation}
T_{{\rm R},{\rm L}} = \frac{1}{1+\lambda^2}
\left[We^{i\phi}+\sum_{ij}G_{ij}(\varepsilon)(We^{i\phi}\Sigma_{ji}+
V_{{\rm L},j}V_{{\rm R},i}^*) \right].
\end{equation}
Finally the conductance is written as
\begin{equation}
G=\frac{2e^2}{h}(2\pi\nu)^2 \left|T_{{\rm R},{\rm L}}\right|^2.
\end{equation}


\begin{thebibliography}{99}
\bibitem{com1}
This is equivalent to eq.\ (4) in ref.\ $[6]$.
\bibitem{com2}
For a single tunnel-junction with tunnel-coupling $W$,
the transmission probability is given by
$[2\lambda/(1+\lambda^2)]^2$.
\bibitem{com3}
In one-dimensional quantum dots, $V_{{\rm L},i}/V_{{\rm R},i}$ of
a level is always out-of-phase with that of the next level.
\end{thebibliography}
\end{document}